\documentclass[aps,prd,twocolumn,showpacs,superscriptaddress,dblfloatfix,preprintnumbers]{revtex4-1}
\usepackage{graphicx}   
\usepackage{dcolumn}   
\usepackage{bm}           
\usepackage{amssymb}  
\usepackage{amsmath}
\usepackage{dsfont}
\usepackage{setspace}
\usepackage{amsmath, amssymb, setspace}
\usepackage{array}
\usepackage{booktabs}
\usepackage{mathrsfs}
\usepackage{indentfirst}
\usepackage{slashed}
\usepackage{float}
\usepackage{lmodern}
\usepackage{hyperref}
\usepackage{xcolor}
\usepackage{multirow}
\usepackage{epstopdf}
\usepackage{ulem}
\usepackage{tabularx}
\usepackage[justification=raggedright]{caption}
\usepackage[flushleft]{threeparttable}
\usepackage{hyperref}
\usepackage{soul}
\usepackage{braket}
\usepackage[T1]{fontenc} 
\usepackage[utf8]{inputenc} 

\DeclareUnicodeCharacter{246}{\"o}	
\DeclareUnicodeCharacter{350}{\c{S}}

\begin{document}


\preprint{IPMU16-0192}

\title{Primordial black holes from scalar field evolution in the early universe}

\author{Eric Cotner}
\affiliation{Department of Physics and Astronomy, University of California, Los Angeles\\
Los Angeles, CA 90095-1547, USA}

\author{Alexander Kusenko}
\affiliation{Department of Physics and Astronomy, University of California, Los Angeles\\
Los Angeles, CA 90095-1547, USA}
\affiliation{Kavli Institute for the Physics and Mathematics of the Universe (WPI), UTIAS\\
The University of Tokyo, Kashiwa, Chiba 277-8583, Japan}

\date{\today}

\begin{abstract}
Scalar condensates with large expectation values can form in the early universe, for example, in theories with supersymmetry. The  condensate can undergo fragmentation into Q-balls before decaying.  If the Q-balls dominate the energy density for some period of time, statistical fluctuations in their number density can lead to formation of primordial black holes (PBH).  In the case of supersymmetry the mass range is limited from above by $10^{23}$g.  For a general charged scalar field, this robust mechanism can generate black holes over a much broader mass range, including the black holes with masses of 1--100 solar masses, which is relevant for LIGO observations of gravitational waves. Topological defects can lead to formation of PBH in a similar fashion.
\end{abstract}


\maketitle

\section{Introduction}

It is well established that stellar core collapse can lead to formation of black holes. However, it remains an open question whether some processes in the early universe could produce primordial black holes (PBH)~\cite{Zeldovich:1967,Hawking:1971ei,Carr:1974nx,GarciaBellido:1996qt,Khlopov:2008qy,Frampton:2010sw,Kawasaki:2012kn,Kawasaki:2016pql,Carr:2016drx,Inomata:2016rbd,Inomata:2017okj,Georg:2017mqk,Carr:2017edp,Domcke:2017fix}. PBHs can account for all or part of dark matter~\cite{Zeldovich:1967,Hawking:1971ei,Carr:1974nx,GarciaBellido:1996qt,Khlopov:2008qy,Frampton:2010sw,Kawasaki:2016pql,Carr:2016drx,Inomata:2016rbd,Inomata:2017okj,Georg:2017mqk,Carr:2017edp,Domcke:2017fix}.  Furthermore, they could be responsible for some of the gravitational wave signals observed by LIGO~\cite{Bird:2016dcv}, In addition, PBHs can invade and destroy neutron stars, ejecting neutron rich material in the process, which can account for for all or part of the 
r-process nucleosynthesis, as well as the 511-keV line in the galactic center~\cite{Fuller:2017uyd}. Finally, PBHs could provide seeds for supermassive black holes~\cite{Kawasaki:2012kn}. A number of scenarios for black hole formation have been considered~\cite{Khlopov:2008qy}, and many of them rely on a spectrum of primordial density perturbations that has some additional power on certain length scales, which can be accomplished by means of tuning an inflaton potential.  

It was recently pointed out that PBHs can form in a very generic scenario, which does not require any particular spectrum of density perturbations from inflation~\cite{Cotner:2016cvr}. Scalar fields with slowly growing potentials form a coherent condensate at the end of inflation \cite{Bunch:1978yq,Linde:1982uu,Affleck:1984fy,Starobinsky:1994bd}. In general, the condensate is not stable, and it breaks up in lumps, which evolve into Q-balls \cite{Kusenko:1997si}. The gas of Q-balls contains a relatively low number of lumps per horizon, and the mass contained in these lumps fluctuates significantly from place to place. This creates relatively large fluctuations of mass density in Q-balls across both subhorizon and superhorizon distances. Since the energy density of a gas of Q-balls redshifts as mass, it can come to dominate the energy density temporarily, until the Q-balls decay, returning the universe to a radiation dominated era. The growth of structure during the Q-ball dominated phase can lead to copious production of primordial black holes.  In this paper we will investigate this scenario in further detail.

Formation of Q-balls requires nothing more than some scalar field with a relatively flat potential at the end of inflation. For example, supersymmetric theories predict the existence of scalar fields with flat potentials. PBH formation in supersymmetric theories is, therefore, likely, even if the scale of supersymmetry breaking exceeds the reach of existing colliders. 

A similar process can occur with topological defects, which can also lead to relatively large inhomogeneities. The discussion of topological defects is complicated by their non-trivial evolution. We will focus primarily on Q-balls, and will briefly comment on topological defects.

The format of this paper is as follows: in Section \ref{sec:FormationOfQballs}, we describe the fragmentation of the condensate and the production of Q-balls, then in Section \ref{sec:ChargeMassDistributions} we derive the formalism for calculating the statistical moments of collections of Q-balls. In Section \ref{sec:PBHdensity} we use the results of the previous Section to calculate the expected PBH density and mass spectrum, and in Section \ref{sec:CosmologicalHistory} we account for the effects on cosmological thermal history and evolve the PBH distribution to the present day. In Section \ref{sec:ExperimentalConstraints}, we then compare our results with current observational constraints, and in Section \ref{sec:ParameterSpace} explore the available parameter space. In Section \ref{sec:TopologicalDefects}, we comment on the applicability of this mechanism to topological defects.

\section{Formation of Q-balls}\label{sec:FormationOfQballs}
Formation of a scalar condensate after inflation and its fragmentation~\cite{Kusenko:1997si} is a fairly generic phenomenon.  While supersymmetry is a well-motivated theory for scalar fields carrying global charges and having flat potentials~\cite{Affleck:1984fy,Dine:2003ax}, our discussion can be easily generalized to an arbitrary scalar field with a global U(1) symmetry in the potential. Supersymmetric potentials generically contain flat directions that are lifted only by supersymmetry breaking terms.  Some of the scalar fields that parameterize the flat directions carry a conserved U(1) quantum number, such as the baryon or lepton number.  During inflation, these field develop a large vacuum expectation value (VEV) \cite{Bunch:1978yq,Linde:1982uu,Affleck:1984fy,Starobinsky:1994bd}, leading to a large, nonzero global charge density. When inflation is over, the scalar condensate $ \phi(t) = \phi_0(t) \exp \{i\theta (t)\}$ relaxes to the minimum of the potential by a coherent classical motion with $\dot\theta\neq 0$ due to the initial conditions and possible CP violation at a high scale.

The initially homogeneous condensate is unstable with respect to fragmentation into non-topological solitons, Q-balls \cite{Coleman:1985ki}. Q-balls exist in the spectrum of every supersymmetric generalization of the Standard Model~\cite{Kusenko:1997zq,Kusenko:1997ad}, and they can be stable or long-lived along a flat direction \cite{Kusenko:1997si,Dvali:1997qv}. In the case of a relatively large charge density (which is necessary for Affleck-Dine baryogenesis \cite{Affleck:1984fy,Dine:2003ax}), the stability of Q-balls can be analyzed analytically \cite{Kusenko:1997si,Enqvist:1997si,Enqvist:1998en}; these results agree well with numerical simulations \cite{Kasuya:1999wu}. One finds that the almost homogeneous condensate develops an instability with wavenumbers in the range $0<k<k_\text{max}$, where $k_\text{max}=\sqrt{\omega^2 - V''(\phi_0)} $, and $\omega=\dot\theta$. The fastest growing modes of instability have a wavelength $\sim 10^{-2\pm 1}$ of the horizon size at the time of fragmentation, and they create isolated lumps of condensate which evolve into Q-balls.  Numerical simulations \cite{Kasuya:1999wu,Kasuya:2000wx} indicate that most of the condensate ends up in lumps. However, since the mass of Q-balls is a non-linear function of the Q-ball size, Q-ball formation, in general, leads to a non-uniform distribution of energy density in the matter component represented by the scalar condensate.  Q-balls can also form when the charge density is small or zero, in which case both positively and negatively charged Q-balls are produced \cite{Kasuya:1999wu}; here we do not consider this possibility. 

Depending on the potential, the Q-balls with a global charge $Q$ have the following properties \cite{Dvali:1997qv,Kusenko:1997si,Kusenko:2005du,Cotner:2016dhw}:
\begin{gather}
\omega \sim \Lambda Q^{\alpha-1}, \qquad R \sim |Q|^\beta/\Lambda, \\
\phi_0 \sim \Lambda |Q|^{1-\alpha}, \qquad M \sim \Lambda |Q|^\alpha,
\end{gather}
where $\Lambda$ is the energy scale associated with the scalar potential, $Q$ is the global U(1) charge and $0<\alpha<1$, $0<\beta<1$ denotes which type of $Q$-ball is under consideration (and also depends on the form of the scalar potential). For ``flat direction" (FD) $Q$-balls, $\alpha = 3/4$ ($\beta = 1/4$), and for ``curved direction" (CD) $Q$-balls, $\alpha = 1$ ($\beta = 1/3$) \cite{Dvali:1997qv,Kusenko:1997si}. 

\section{Q-ball charge/mass distributions}\label{sec:ChargeMassDistributions}
Numerical simulations of condensate fragmentation and $Q$-ball formation have been performed in the past, from which we are able to determine the resulting charge and mass distributions \cite{Multamaki:2002hv,Kasuya:2000wx}. These distributions appear to be very sensitive to initial conditions in the condensate, such as the ratio of energy to charge density ($x=\rho/mq$), and to the details of the scalar potential. In addition, the resultant charge distribution can be very non-Gaussian due to the high degree of nonlinearity and chaos in the fragmentation process.

\begin{figure}
\centering
\includegraphics[width=0.9\linewidth]{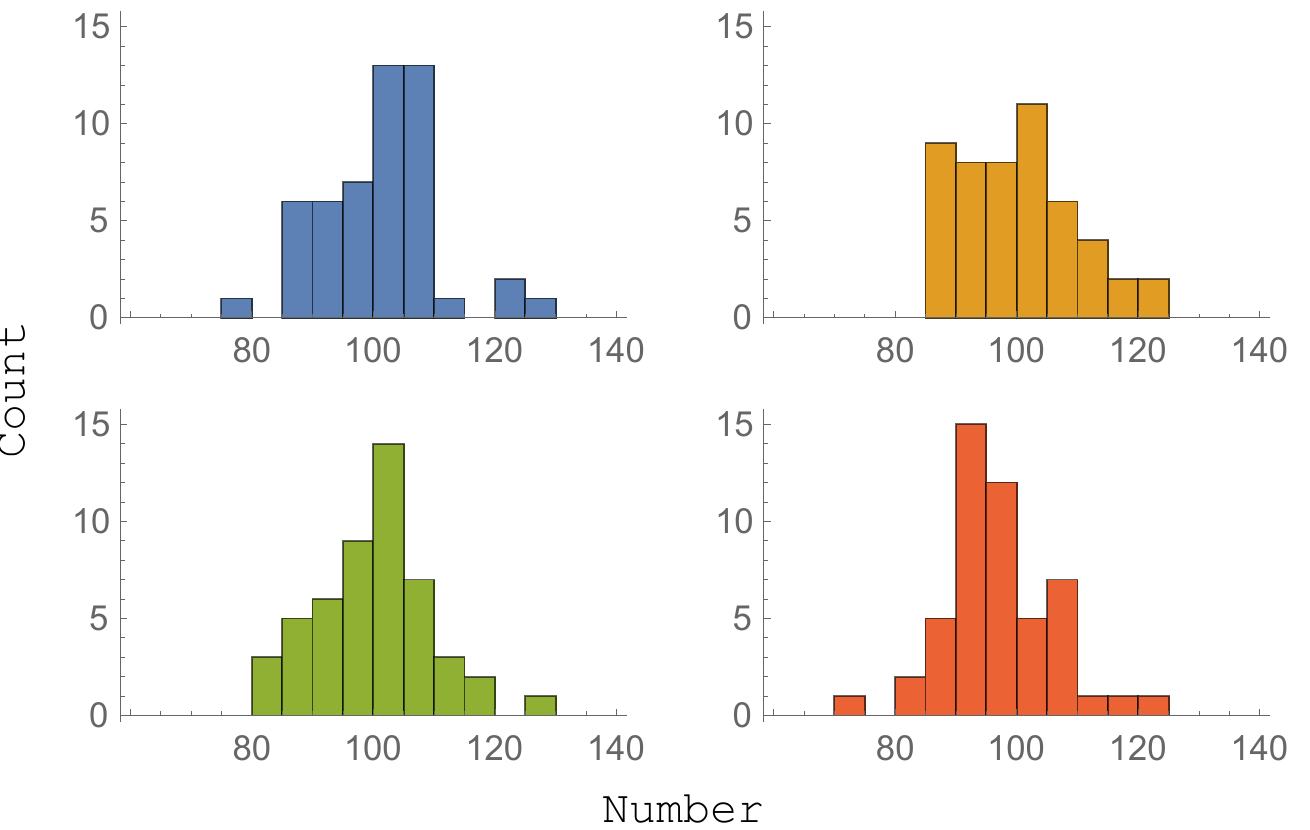}
\caption{Schematic illustration of 4 histograms each containing 100 samples from the same Poisson distribution ($\lambda = 100$). The differences due to fluctuations are clearly visible.}
\label{fig:LowSampleNumber}
\end{figure}

It should be understood that the results of these simulations are statistical in nature: a large number of $Q$-balls are created within the simulation volume so that the charge distribution tends towards a statistical average. In reality, if one were to perform a large simulation and look at the charge distributions in a number of small sub-volumes, you will find a large degree of variation, with more variation on smaller scales due to small sample sizes (this is not to say that the \textit{variance} will be larger, just that the differences between distributions are large), as can be seen in Fig. \ref{fig:LowSampleNumber}.
It is these large fluctuations relative to the mean that will be the source of density perturbations.


Once the resultant charge distribution of Q-balls $f_Q(Q)\, dQ$ has been calculated from these numerical simulations, we can use this to calculate the mass distribution for single Q-balls using $M=\Lambda|Q|^\alpha$ (we will absorb all numerical factors into the definition of $\Lambda$ without loss of generality):
\begin{gather}
f_M(M) = \frac{M^\frac{1-\alpha}{\alpha}}{\alpha \Lambda^{1/\alpha}} \left[f_Q((M/\Lambda)^{1/\alpha}) + f_Q(-(M/\Lambda)^{1/\alpha})\right].
\end{gather}
It is important to note that a distribution well-localized in charge is also well-localized in mass. We can also use probability theory to calculate the mass of a collection of Q-balls. Under the assumption that a charge $Q_\text{tot}$ is distributed amongst $N$ Q-balls whose distribution is described by $f_Q(Q)$, the probability distribution function (PDF) for the total mass of this collection of Q-balls is given by
\begin{gather}
f_M(M,Q_\text{tot},N) = \frac{\psi(M,Q_\text{tot},N)}{\int dM\, \psi(M,Q_\text{tot},N)}, \label{eq:MassDistribution} \\
\begin{align}
\psi(M,Q_\text{tot},N) &= \int \left(\prod_{i=1}^{N} dQ_i\, f_Q(Q_i)\right) \\
&\times \delta\left(M - \Lambda \sum_{i=1}^{N} |Q_i|^\alpha\right) \delta\left(Q_\text{tot} - \sum_{i=1}^{N} Q_i\right), \nonumber
\end{align}
\end{gather}
where $\psi$ admits a simple-looking Fourier transform in $M$ and $Q_\text{tot}$:
\begin{gather}
\tilde{\psi}(\xi_M,\xi_{Q_\text{tot}},N) = \left[ \int dQ\, e^{i(\xi_M |Q|^\alpha + \xi_{Q_\text{tot} Q})} f_Q(Q) \right]^N.
\end{gather}
The power of $\alpha$ prevents analytic calculation of this PDF for all but the simplest charge distributions. Specifically, if we take the charge distribution to be a delta function: $f_Q(Q) = \delta(Q - Q_0)$, then the mass distribution is also a delta function: $f_M(M,Q_\text{tot},N) = \delta(M - N M_0)$, where $M_0 = \Lambda Q_0^\alpha$, and $Q_0 = Q_\text{tot}/N$ to satisfy charge conservation (this constraint comes from a mathematical issue that arises due to the canceling of a delta function of the form $\delta(Q_\text{tot}-NQ_0)/\delta(Q_\text{tot}-NQ_0)$; we can see that if we consider $\delta(x)$ as the limit of a smooth function that approaches this distribution, then this ratio is unity provided $Q_0 = Q_\text{tot}/N$).

For ease of computation, we will assume the delta function charge/mass distribution for the rest of this paper. This also has good theoretical motivation, as the Affleck-Dine baryogenesis scenario requires a large nonzero charge density, which tends to result in a highly-localized charge distribution.

\subsection{Single length scale}
One should notice that the mass distribution function calculated earlier is also a function of both the total charge $Q_\text{tot}$ and the number of Q-balls $N$. During the chaotic fragmentation procedure, the number of Q-balls will fluctuate between horizons. So in order to get a full description of the fluctuations, we must supplement the mass distribution with a number distribution $p(N)$. This can be calculated from a simulation by simply counting the number of Q-balls within the simulation volume. Here, we will assume that the number of Q-balls per horizon $N$ is described by a Poisson distribution, as is typical for a random process such as fragmentation:
\begin{gather} \label{eq:HorizonNumberDistribution}
p(N) = e^{-N_f} \frac{N_f^N}{N!},
\end{gather}
where $N_f$ is the average number of Q-balls per horizon at fragmentation. We then combine Equations \ref{eq:MassDistribution} and \ref{eq:HorizonNumberDistribution} to create a joint PDF which describes the distribution of mass $M$ within a horizon composed of $N$ Q-balls (we also set $Q_\text{tot} = Q_f$, the total charge on the horizon at $t_f$): $F_Q(M,N) = f_M(M,Q_f,N) p(N)$. This is manifestly normalized since $\sum_N \int dM\, F_Q = \sum_N (\int dM\, f_M) p = \sum_N p = 1$. We can then use this to calculate statistical moments such as the average horizon mass, average horizon Q-ball number, RMS fluctuations, etc:
\begin{gather}
\braket{M} = \sum_{N=1}^{\infty} \int_{0}^{\infty} dM\, M F_Q(M,Q_f,N) \approx M_f, \\
\braket{N} = \sum_{N=1}^{\infty} \int_{0}^{\infty} dM\, N F_Q(M,Q_f,N) = N_f, \\
N_\text{RMS} = \left[\sum_{N=1}^{\infty} \int_{0}^{\infty} dM\, N^2 F_Q(M,Q_f,N)\right]^{1/2} = N_f,
\end{gather}
where $M_f = \Lambda Q_f^\alpha N_f^{1-\alpha}$ is the horizon mass of Q-balls at $t_f$ (the first relation is only approximate because $\braket{N^{1-\alpha}} \approx N_f^{1-\alpha}$, though the relative error scales as $|\braket{N^{1-\alpha}} - N_f^{1-\alpha}|/\braket{N^{1-\alpha}} \approx 1/10 N_f$, so totally negligible for large $N_f$).

\subsection{Multiple length scales}\label{ssec:MultipleLengthScales}
The previous treatment has the shortcoming that it can only describe Q-ball distributions with spatial extent the size of the horizon at the time of fragmentation. We now generalize this to handle distributions on an arbitrary scale. First, when considering a physical volume $V$ at the time of fragmentation, the charge contained within this volume (assuming initial uniformity of the condensate) is given by $Q_\text{tot} = Q_V = Q_f (V/V_f)$, where $V_f = \frac{4\pi}{3} t_f^3$ is the horizon volume at $t_f$. Second, the number distribution is altered so that the number of Q-balls within volume $V$ (assuming the same average number density $n_f = N_f/V_f$ across all scales) is described by
\begin{gather}\label{eq:NumberDistribution}
p(N,V) = e^{-N_f V/V_f} \frac{(N_f V/V_f)^N}{N!}.
\end{gather}
The joint PDF for a mass $M$ composed of $N$ Q-balls contained within a volume $V$ at the time of fragmentation $t_f$ is then given by
\begin{gather}\label{eq:QballClusterMassDistribution}
F_Q(M,V,N) = \delta\left(M - M_f \left(\frac{N}{N_f}\right)^{1-\alpha} \left(\frac{V}{V_f}\right)^\alpha\right) p(N,V).
\end{gather}
Note that if $V=V_f$, this reduces to the single-scale, horizon-size treatment.

In addition to being able to calculate quantities on each scale $V$ individually, we will also want to sum the contributions from each scale in some cases (such as contributions to the PBH density from both subhorizon and superhorizon modes). To do so, we will sum over all volume scales from $V_\text{min}$ to $V_\text{max}$ using a coarse-graining method. We consider an arbitrary function of volume $g(V)$. The sum of the contributions from each scale $V_i = V_\text{max}/\chi^{i-1}$ is then given by
\begin{align}
\sum_{\{V\}} g(V) &= \sum_{i=1}^{\lfloor 1 + \log_\chi\frac{V_\text{max}}{V_\text{min}} \rfloor} g(V_i) \\
&\approx \int_{1}^{ 1 + \log_\chi\frac{V_\text{max}}{V_\text{min}}} di\, g(V_\text{max}/\chi^{i-1}) \\
&= \frac{1}{\ln\chi} \int_{V_\text{min}}^{V_\text{max}} \frac{dV}{V}\, g(V),
\end{align}
where we have used Euler-Maclaurin to approximate the sum, and $\chi \sim \text{few}$ is a parameter of the spacing between intervals of the coarse-graining procedure. We will take $\chi = e$ from now on for simplicity; another choice does not significantly affect the outcome provided it is not too close to unity. $V_\text{min}$ is the smallest volume under consideration; there will be a natural cutoff due to the fact that Q-balls have a finite size, and so this scale is generally defined as the volume which contains some number $N_\text{min} \sim 10$ Q-balls on average: $V_\text{min}/N_\text{min} = N_f/V_f$.

\section{Q-ball and PBH densities}\label{sec:PBHdensity}
Using the framework of Section \ref{sec:ChargeMassDistributions}, we are now in a position to begin calculating the energy densities associated with the Q-balls, fluctuations in that energy density, and the resulting density of black holes.

\subsection{Q-ball density at fragmentation}
Using the formalism of Section \ref{ssec:MultipleLengthScales}, we can calculate the background energy density (over the largest scales) of Q-balls at $t_f$:
 \begin{gather}\label{eq:QballDensity}
\braket{\rho_Q(t_f)} = \lim\limits_{V\rightarrow\infty} \frac{\braket{M}}{V} = \frac{M_f}{V_f},
\end{gather}
which will be important in the discussion of density perturbations in Section \ref{sec:PBHdensity}. Since Q-balls are formed at rest, the evolution of the Q-ball density after fragmentation is simply that of decaying nonrelativistic matter $\braket{\rho_Q(t)} = \braket{\rho_Q(t_f)} (a_f/a)^3 e^{(t_f-t)/\tau_Q}$, where $\tau_Q = 1/\Gamma_Q$ is the lifetime of the Q-balls. Q-balls are generally considered stable with respect to decay into the quanta of the scalar field, but it is possible to decay through other processes. For example, if a coupling of the scalar field to a light fermion with mass $m<\omega$ exists, Q-balls can decay to these fermions through an evaporation process \cite{Cohen:1986ct,Kawasaki13,Enqvist:1998xd}. Q-balls can also decay if the U(1) symmetry is broken by some higher-dimension operators \cite{Kusenko:2005du,Kawasaki:2005xc,Kasuya:2014ofa,Cotner:2016dhw}. We define $\Gamma_Q$ to include all such decay channels.

\subsection{Q-ball density perturbations due to fluctuations}\label{ssec:QballDensityPerturbations}
Due to the stochastic nature of the fragmentation process, volumes of space can arise within which the number density of Q-balls exceeds the average number density. Due to the nonlinear relationship between Q-ball mass and charge $M=\Lambda|Q|^\alpha$, this also gives rise to fluctuations in the energy density within that volume. The density contrast in Q-balls at fragmentation $\delta(t_f)$ for a volume $V$ containing mass $M$ is defined as
\begin{gather}
\delta(t_f) = \frac{\delta\rho_Q}{\braket{\rho_Q}} = \frac{M/V}{\braket{\rho_Q}} - 1 = \left(\frac{N/N_f}{M/M_f}\right)^\frac{1-\alpha}{\alpha} - 1
\end{gather}
where in the last line we have used the argument of the delta function in Equation \ref{eq:QballClusterMassDistribution} to eliminate $V$ (this will be justified by an integral over $V$ later). Note that if the Q-ball mass-charge relationship were linear ($\alpha=1$), the perturbations would vanish identically.

The subhorizon density perturbations ($V<V_f$) are frozen during the initial radiation dominated era, but they grow linearly in the scale factor during the Q-ball dominated epoch: $\delta(t) = \delta(t_f) (a/a_Q) = \delta(t_f) = (t/t_Q)^{2/3}$, where $t_Q$ is the beginning of the era of Q-ball domination. The structure growth generally goes nonlinear and decouples from the expansion around $\delta > \delta_c \sim 1.7$, at which point the overdense regions collapse and become gravitationally bound. However, some structures with $\delta < \delta_c$ can still collapse, and not all structures with $\delta > \delta_c$ are guaranteed to collapse into black holes. Due to nonsphericity of the gravitationally-bound structures, only a fraction $\beta = \gamma \delta^{13/2}(t_R)(M/M_Q)^{13/3}$ (where $\gamma \approx 0.02$ is a factor due to the nonsphericity, $M_Q = M_f(t_Q/t_f)^{3/2}$ is the horizon mass at the beginning of the Q-ball dominated era, and $t_R$ is the end of the Q-ball dominated era, when the radiation comes to dominate again) will actually collapse to black holes \cite{Carr:2016drx,Polnarev:1986bi,Georg:2016yxa} by the end of the Q-ball dominated era. We assume that structures with $\delta \ge \delta_c$ do not continue to grow past the point of nonlinearity, as they have already collapsed and had their chance to form a PBH; for these perturbations we set $\beta = \gamma \delta_c^{13/2} (M/M_Q)^{13/3}$ for $\delta(t_R) > \delta_c$. This refinement may not be necessary, as the average density perturbations are generally so small they never reach $\delta_c$, and indeed, changing the value of $\delta_c$ does not seem to significantly alter the outcome.

Additional care must be taken to extend this to scales which enter the horizon at later times, and thus may not subject to the same amount of growth as subhorizon modes. Those that enter the horizon between $t_f < t < t_Q$ can be treated as effectively subhorizon since they enter the horizon before the Q-ball dominated epoch begins, and thus fluctuations are subject to the same amount of amplification as initially subhorizon modes. This includes all volumes $V < V_Q$, where $V_Q = \frac{4\pi}{3} t_Q^3 (t_f/t_Q)^{3/2}$ is defined as the (initially superhorizon) physical volume at $t_f$ which enters the horizon at $t_Q$: $(a_Q/a_f)^3 V_Q = \frac{4\pi}{3} t_Q^3$. Fluctuations which enter the horizon during the Q-ball dominated epoch are treated slightly differently, as they are only subjected to amplification from the time they enter the horizon $t_h$ until the radiation comes to dominate again at $t_R$. We can account for this by calculating $t_h$ for a given scale $V$ via $(a(t_h)/a_f)^3 V = \frac{4\pi}{3} t_h^3$ (which gives us $t_h = (3/4\pi)(V/t_f^{3/2} t_Q^{1/2})$), and then replacing the scale factor $a_R/a_Q$ with $a_R/a(t_h)$ in the definition of $\beta$ above. This treatment is valid for all scales between $V_Q < V < V_R$, where $V_R = (4\pi/3) t_R^3 (t_Q/t_R)^2 (t_f/t_Q)^{3/2}$ is the physical volume at $t_f$ which enters the horizon at $t_R$.

In addition to these details, we also enforce the constraint $\beta \le 1$ in order to prevent PBH production probabilities over unity, though this does not become relevant unless the Q-ball dominated era is extremely long.

\subsection{Primordial black hole density}
We are now in a position to calculate the average energy density in PBH created during the Q-ball dominated era. We do this by calculating the energy density of Q-balls at $t_f$ that \textit{will eventually} form black holes by $t_R$ by weighting the Q-ball energy density $M/V$ by the collapse fraction/probability $\beta$ evaluated at $t_R$, summing over all scales $V$, and then redshifting this value appropriately. The expression for this procedure is given by
\begin{gather}\label{eq:BHdensity}
\braket{\rho_\text{BH}(t_R)} = \left(\frac{a_f}{a_R}\right)^3 \sum_{N=1}^{\infty} \int_{V_\text{min}}^{V_R} \frac{dV}{V} \int_{0}^{\infty} dM\, \left(\beta \frac{M}{V}\right) F_Q,
\end{gather}
where it should be understood that the integral over $V$ is broken up into two separate domains, $[V_\text{min},V_Q]$ and $[V_Q,V_R]$, where separate definitions of $\beta$ apply, as described in Section \ref{ssec:QballDensityPerturbations}. Due to the complicated piecewise nature of the function $\beta$, the authors are unaware of any analytic solution, and further progress must be made numerically.

\begin{figure}
\centering
\includegraphics[width=0.9\linewidth]{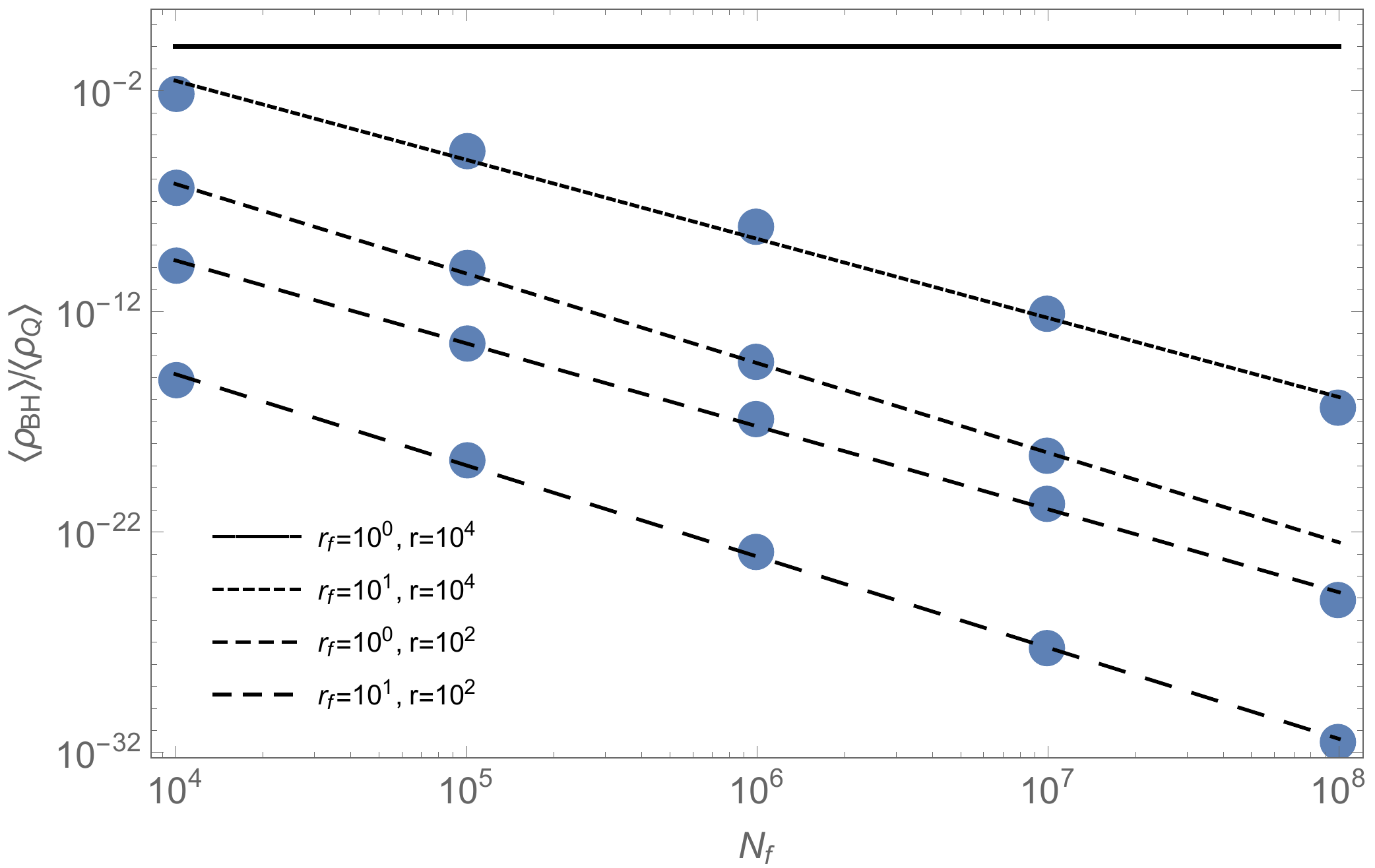}
\caption{Fraction of Q-ball energy that goes into black holes. The ratio roughly scales as $\braket{\rho_\text{BH}}/\braket{\rho_Q} \sim (4.6\times 10^{-4}) r_f^{-5.9} r^{4.4} N_f^{-3.7}$. The thick black line is the unity bound where 100\% of the Q-ball energy goes into creating black holes.}
\label{fig:DensityRatioVsNf}
\end{figure}

We find that Equation \ref{eq:BHdensity} can be rewritten in such a way that it only depends on the dimensionless numbers $N_f$, $r_f = t_Q/t_f$, and $r = t_R/t_Q$. $r_f$ and $r$ can be interpreted as measures of the duration of the era between the fragmentation and the beginning of Q-ball domination, and the length of the Q-ball dominated era, respectively. The effect of these parameters on the black hole density can be seen in Figure \ref{fig:DensityRatioVsNf}.
Larger $r_f$ will reduce the fraction of Q-ball energy that goes into making black holes due to the dilution of the number density and increased horizon mass at $t_Q$ due to the delay of the Q-ball dominated era. Larger $r$ leads to an increased fraction of Q-ball energy that goes into black holes due to more amplification of the density perturbations, leading to a higher probability of PBH formation. Larger $N_f$ reduces the fraction because of higher suppression of fluctuations due to the Poisson statistics. The form of the contours in this plot suggest that this ratio roughly scales as $\braket{\rho_\text{BH}}/\braket{\rho_Q} \sim (4.6\times 10^{-4}) r_f^{-5.9} r^{4.4} N_f^{-3.7}$.

\subsection{Black hole mass spectrum}
\begin{figure}
\centering
\includegraphics[width=0.9\linewidth]{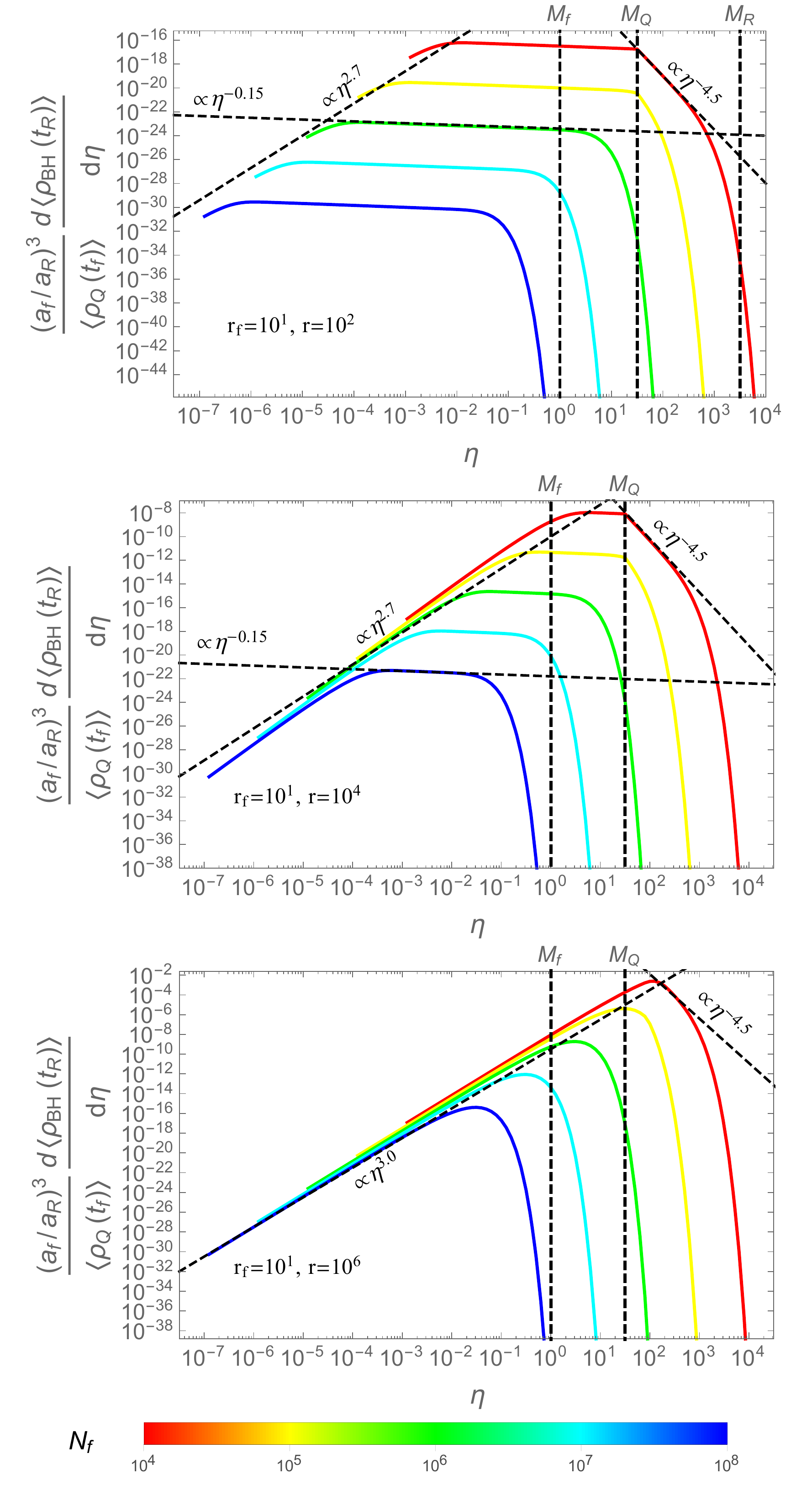}
\caption{Differential PBH/Q-ball density ratio as a function of $\eta=M/M_f$. The density of black holes at $t_R$ has been blueshifted back to $t_f$ for comparison with the initial Q-ball density. The integral of these curves over $\eta$ is the fraction of energy in Q-balls that goes into black holes, as shown in Figure \ref{fig:DensityRatioVsNf}. Notice that as $r$ increases, the flat region shrinks. Parameters for the given spectra are shown in the lower left corner.}
\label{fig:MassSpectra}
\end{figure}

One can derive the mass spectrum of the black holes by not integrating over $M$ in Equation \ref{eq:BHdensity}:
\begin{gather}\label{eq:BHspectrum}
\frac{d\braket{\rho_\text{BH}}}{dM} = \left(\frac{a_f}{a_R}\right)^3 \sum_{N=1}^{\infty} \int_{V_\text{min}}^{V_R} \frac{dV}{V} \left(\beta \frac{M}{V}\right) F_Q.
\end{gather}
This yields the differential black hole energy density $d\braket{\rho_\text{BH}}/dM$. We find that the spectrum can be rewritten in terms of the parameter $\eta = M/M_f$ (fraction of horizon mass at $t_f$), along with the previously mentioned parameters $r_f=t_Q/t_f$, $r=t_R/t_Q$, and $N_f$. Calculation of this function can be done by evaluating Equation \ref{eq:BHspectrum} at multiple values of $\eta$ and then interpolating. An example is given in Figure \ref{fig:MassSpectra}.
First, it's obvious from the normalization of each curve that the lower the number of Q-balls per horizon, the more black holes that are created. This is expected, as the Poisson statistics suppress the density fluctuations for large Q-ball number. The normalization also increases with $r$, as explained in Figure \ref{fig:DensityRatioVsNf}. Second, there is a hard lower cutoff in the PBH mass, which occurs at $\eta = N_\text{min}/N_f$, which is due to the lower cutoff in the volume mentioned earlier. Above that, the BH number sharply increases with a power law $\propto \eta^{2.85\pm 0.15}$; the extent of this region depends on the magnitude of $r$, with larger values leading to a larger range. We suspect that this is due to the fact that the small-scale density fluctuations have already reached their critical value $\delta_c$ and can no longer continue growing, whereas the large-scale fluctuations (which started out smaller) still have room to do so. Above that, the spectrum becomes approximately flat ($\propto \eta^{-0.15}$), meaning that the number of black holes in each decade of mass are comparable. Of course, the upper end of this range dominates the energy density of the distribution. Then, at around $M = M_Q$, there is a sharp transition and the slope becomes strongly negative ($\propto \eta^{-4.5}$) due to the reduced growth the superhorizon modes are subject to. Then, there is an upper exponential cutoff at $\eta \sim 10^8/N_f$ due once again to the Poisson statistics (the cutoff appears to take precedence over previously mentioned transitions).

\section{Cosmological history}\label{sec:CosmologicalHistory}
\begin{figure}
\centering
\includegraphics[width=0.9\linewidth]{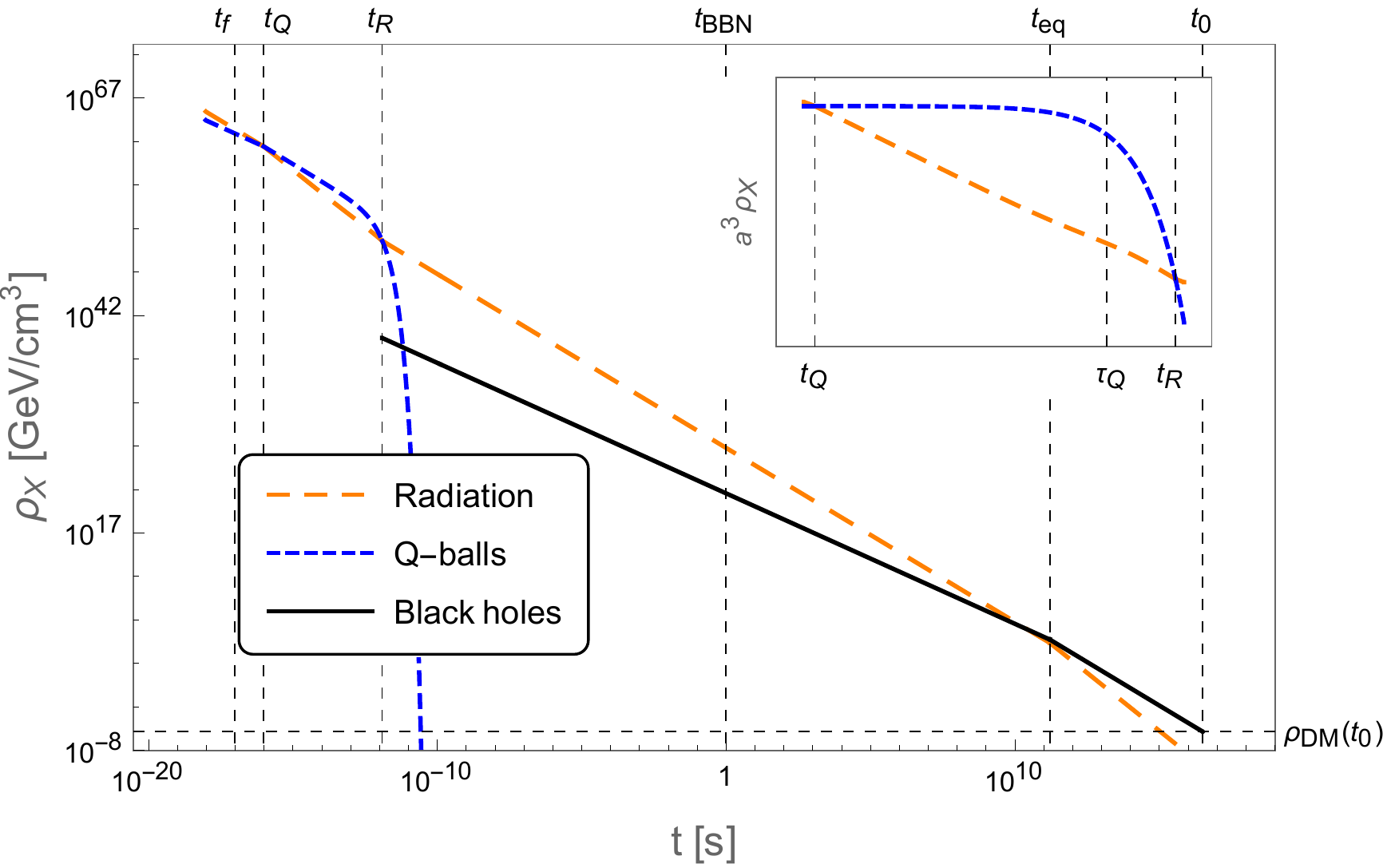}
\caption{Evolution of energy density over cosmological history. The evolution of three ``species" is plotted: radiation $\rho_R$ (orange, long dash), Q-balls $\braket{\rho_Q}$ (blue, short dash), and black holes $\braket{\rho_\text{BH}}$ (black, solid). Inset in the upper right corner is a zoomed-in view of the Q-ball dominated era ($t_Q < t <t_R$). The vertical axis is scaled by a factor of $a^3$, so that the decay of the Q-balls is evident (non-decaying matter would be represented by a straight horizontal line). The parameters of this model are $N_f = 10^6$, $t_f = 9.7\times 10^{-18}\text{ s}$, $r_f = 10$, and $r=1.3\times 10^4$, which corresponds to production of PBH with peak mass of $4.4\times 10^{20}\text{ g}$ making up 100\% of the dark matter.}
\label{fig:CosmologicalTimeline}
\end{figure}

We now give a detailed account of how the Q-balls, radiation, and black holes evolve throughout the history of the universe up until the present day, as seen in Figure \ref{fig:CosmologicalTimeline}.
In summary: we assume an initial period of inflation and reheating in order to create a radiation dominated era with a uniform charged scalar field as a subdominant component of the energy density. The scalar field fragments into Q-balls at $t_f$, which then come to dominate the energy density at $t_Q$. During the Q-ball dominated epoch, primordial black holes are produced, and at $t_R$, their density is frozen in and evolves as nonrelativistic matter. After this initial matter dominated epoch, the Standard Model of cosmology resumes, and evolves through all the eras we are familiar with (BBN, matter-radiation equality, etc.) up to the present day.

The functions used to model the energy density evolution for each species is summarized in Appendix \ref{sec:EvolutionOfEnergyDensities}.

\subsection{Initial radiation dominated era}
After the end of inflation, the Universe enters a brief matter dominated era due to the coherent oscillations of the inflaton field. The decay of the quanta of this field at time $t_\text{RH} = \Gamma_I^{-1}$ reheats the Universe, which enters a radiation dominated epoch with temperature $T_\text{RH} = 0.55 g_*^{-1/4} (\Gamma_I M_p)^{1/2}$ and radiation energy density
\begin{gather}
\rho_R(t_\text{RH}) = \frac{\pi^2}{30} g_*(T_\text{RH}) T_\text{RH}^4 \approx \frac{\pi^2}{327} \Gamma_I^2 M_p^2,
\end{gather}
where $\Gamma_I \sim 1/t_\text{RH}$ is the decay width of the inflaton oscillations. The radiation density redshifts as $\rho_R(t) = \rho_R(t_\text{RH}) (a_\text{RH}/a)^4 = \rho_R(t_\text{RH}) (t_\text{RH}/t)^2$ during this epoch, which ends up canceling the factor of $\Gamma_I$ to give us
\begin{gather}\label{eq:RadDensityRadDominated}
\rho_R(t) = \frac{\pi^2 M_p^2}{327 t^2};\qquad t_\text{RH} < t < t_Q
\end{gather}
At some point $t_f$, the scalar condensate fragments into Q-balls, resulting in an energy density given by Equation \ref{eq:QballDensity}. The Q-balls then redshift as decaying nonrelativistic matter:
\begin{align}\label{eq:QballDensityRadDominated}
\braket{\rho_Q(t)} &= \braket{\rho_Q(t_f)} \left(\frac{a_f}{a}\right)^3 e^{-(t-t_f)/\tau_Q} \\
&= \frac{3 \Lambda Q_f^\alpha N_f^{1-\alpha}}{4\pi t_f^{3/2} t^{3/2}} e^{-(t-t_f)/\tau_Q} ;\qquad t_f < t < t_Q
\end{align}

\subsection{Q-ball dominated era}
At some point $t_Q$, the Q-balls come to dominate the energy density. This time is defined by $\rho_R(t_Q) = \braket{\rho_Q(t_Q)}$, using the equations of the previous section. During this era, Q-ball decays begin to affect the radiation density, causing the radiation temperature to decrease less slowly than it normally would due to the expansion. Following the analysis of Scherrer and Turner \cite{Scherrer:1984fd}, the radiation density in this epoch due to the decay of the Q-balls can be modeled as
\begin{gather}\label{eq:RadDensityQballDominated}
\rho_R(t) = \left[ \rho_R(t_Q) + \braket{\rho_Q(t_Q)} \int_{x_0}^{x} dx'\, z(x') e^{-x'} \right] z^{-4},
\end{gather}
where $x \equiv \Gamma_Q t$, $x_0 = \Gamma_Q t_Q$, and $z=(x/x_0)^{2/3}$. The Q-balls continue to redshift and decay, leading to
\begin{gather}\label{eq:QballDensityQballDominated}
\braket{\rho_Q(t)} = \frac{3 \Lambda Q_f^\alpha N_f^{1-\alpha} t_Q^{1/2}}{4\pi t_f^{3/2} t^2} e^{-(t-t_f)/\tau_Q} ;\qquad t_Q < t < t_R
\end{gather}
As the Q-balls decay, eventually the radiation comes to dominate again at $t_R$, defined by $\rho_R(t_R) = \braket{\rho_Q(t_R)}$. Using Equations \ref{eq:RadDensityRadDominated}, \ref{eq:QballDensityRadDominated}, \ref{eq:RadDensityQballDominated} and \ref{eq:QballDensityQballDominated}, this gives us the relation
\begin{gather}
1 + \left(\frac{t_R}{\tau_Q}\right)^{-2/3} \Gamma\left(\frac{5}{3}, \frac{t_Q}{\tau_Q}, \frac{t_R}{\tau_Q}\right) = \left(\frac{t_R}{t_Q}\right)^{2/3} e^{(t_Q - t_R)/\tau_Q},
\end{gather}
where $\Gamma$ is the generalized incomplete gamma function. This allows us to solve (numerically) for $r_Q \equiv \tau_Q/t_Q$ as a function of $r=t_R/t_Q$. At this point, if we specify $t_f$, $r_f$ and $r$, we can calculate the other parameters $t_Q$, $t_R$, $\tau_Q$, and $\Lambda Q_f^\alpha$ via
\begin{gather}
t_Q = t_f r_f, \quad t_R = t_f r_f r,\quad \tau_Q = t_f r_f r_Q(r), \\
\Lambda Q_f^\alpha = \frac{4\pi M_p^2 t_f}{3\cdot 327 r_f^{1/2} N_f^{1-\alpha}} e^{(1-1/r_f)/r_Q(r)},
\end{gather}
from which we can calculate all other quantities of interest ($M_f = \Lambda Q_f^\alpha N_f^{1-\alpha}$, $M_Q = M_f r_f^{3/2}$, etc.).

\subsection{Standard cosmological era}\label{ssec:StandardCosmology}
After the Q-balls have decayed sufficiently, the universe returns to a radiation dominated era, and the standard cosmology begins. In order to evolve the radiation, Q-ball, and black hole densities to the present day, one would na\"ively use $a_1/a_2 = (t_1/t_2)^n$, where $n=\frac{1}{2}\,(\frac{2}{3})$ in a radiation (matter) dominated era, keeping in mind that the universe transitions between the two at $z_\text{eq} \approx 3360$, or $t_\text{eq} \approx 4.7\times 10^4\text{ yr}$. However, due to the extended era of matter domination, the time at which cosmological events (such as BBN, matter-radiation equality, or recombination) occur are not the same as in the standard cosmology. Instead, one must evolve according to the universe's \textit{thermal} history, where cosmological events occur at specific temperatures. In this case, one must use $a_1/a_2 = g_{*S}^{1/3}(T_2) T_2/g_{*S}^{1/3}(T_1) T_1$ and evolve from $T_R$ (defined by $\rho_R(t_R) = (\pi^2/30) g_{*}(T_R) T_R^4$) to $T_0 = 2.7 \text{ K} = 2.3 \text{ meV}$. This has the advantage of accurately accounting for any deviation from cosmological history. We can then find the time at which some event $X$ occurs by solving $\rho_R(t_X) = (\pi^2/30) g_{*}(T_X) T_X^4 = \rho_R(T_R) (a(t_R)/a(t))^4$. In order to ensure that this early matter dominated era does not spoil the canonical cosmological thermal history, we enforce an additional constraint $T_R > T_\text{BBN} \sim \text{MeV}$, so that the entropy injection from Q-ball decays does not interfere with nucleosynthesis.

\section{Observational constraints}\label{sec:ExperimentalConstraints}
\begin{figure}
\centering
\includegraphics[width=0.9\linewidth]{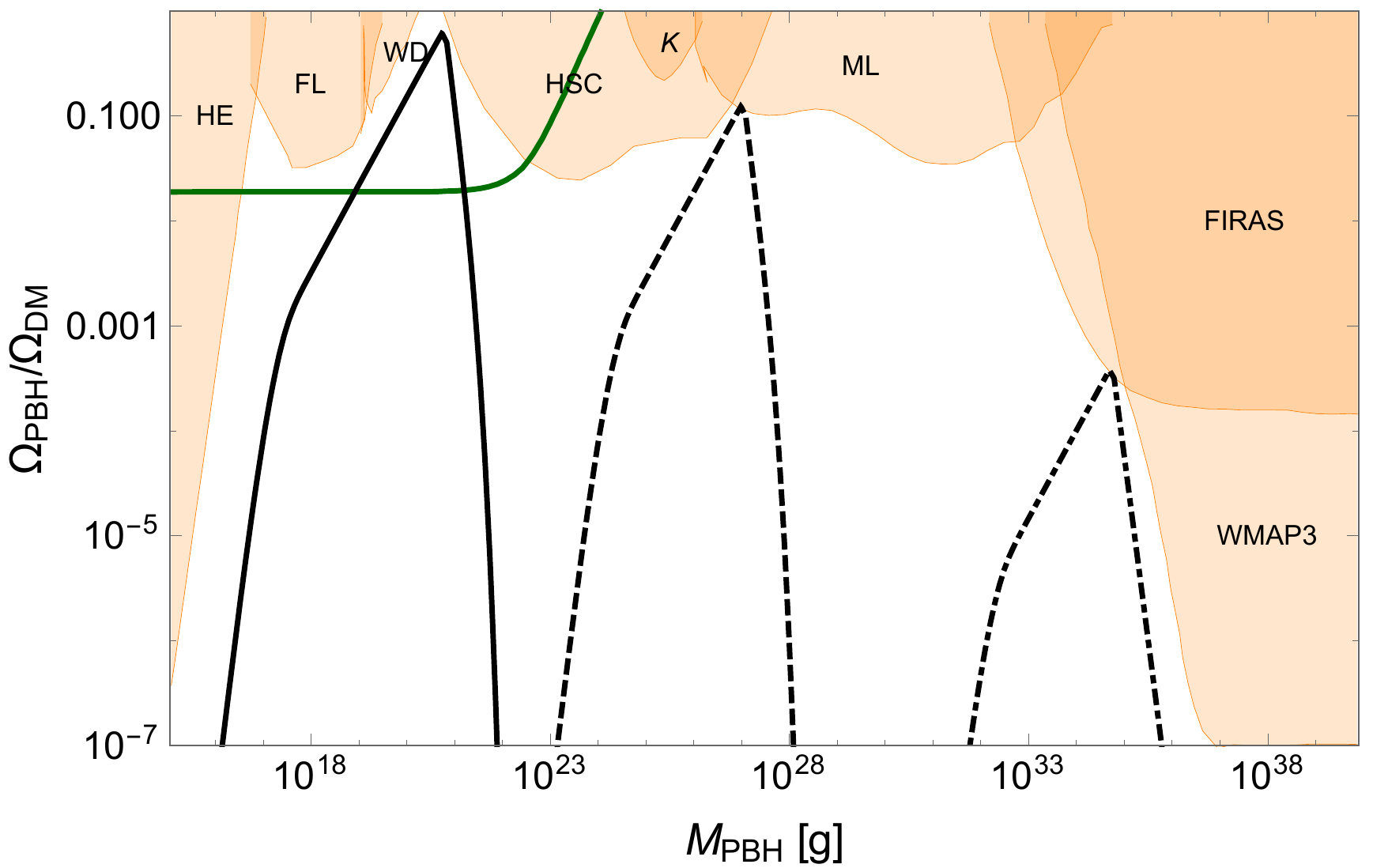}
\caption{Comparison of observational PBH constraints $f_\text{con}(M) = \Omega_\text{PBH}(M)/\Omega_\text{DM}$ (orange, shaded) with the dark matter fraction per logarithmic interval (black), defined by $f(M) = \frac{M}{\rho_\text{DM}} \frac{d\braket{\rho_\text{BH}(t_0)}}{dM}$. This is a crude comparison, and to be rigorous one should use the procedure outlined by the use of Equation \ref{eq:DMfractionBin}. Parameters for the three curves are $t_f = 1.12\times 10^{-17}\text{ s}$, $r_f=1.1$, $r=4.47\times 10^2$, $N_f=10^6$, $f=1$ (solid line), $t_f=2.0\times 10^{-11} \text{ s}$, $r_f = 1.1$, $r=1.58\times 10^3$, $N_f = 10^6$, $f=0.2$ (dashed line), and $t_f=1.0\times 10^{-3}\text{ s}$, $r_f=1.1$, $r=4.47\times 10^2$, $N_f=10^5$, $f=0.001$ (dot-dashed line). Also shown is the boundary of the parameter range (dark green) above which PBH collisions with neutron stars can account for all r-process  element production in the
Milky Way \cite{Fuller:2017uyd}.}
\label{fig:CombinedConstraintPlots}
\end{figure}

We now examine the observational constraints on primordial black holes and where our results fit in. The constraints come from a wide variety of sources, such as extragalactic gamma rays from evaporation \cite{2017APS..APR.K6004J,PhysRevD.94.044029}, femtolensing of gamma ray bursts (GRB) \cite{Barnacka:2012bm}, capture by white dwarfs \cite{PhysRevD.92.063007}, microlensing observations from HSC \cite{Niikura:2017zjd}, Kepler \cite{Griest:2013esa,Griest:2013aaa}, and EROS/MACHO/OGLE \cite{Tisserand:2006zx}, measurements of distortion of the CMB \cite{PhysRevD.78.023004,Ricotti:2007au}, and bounds on the number density of compact X-ray objects \cite{Inoue:2017csr} (constraints summarized in \cite{Carr:2009jm,Frampton:2010sw,Kuhnel:2017pwq}). The constraints are typically expressed in a form that assumes a monochromatic distribution of PBH masses. However, in the case of an extended mass distribution (such as we have in this scenario), care must be taken to apply the limits correctly. To do this, we follow the procedure outlined in \cite{Carr:2016drx,Kuhnel:2017pwq}, which amounts to dividing the mass spectrum into a number of bins (labeled by the index $i$), then integrating the dark matter fraction over the interval contained in the bin:
\begin{gather}\label{eq:DMfractionBin}
f_i = \frac{1}{\rho_\text{DM}} \int_{M_i}^{M_{i+1}} dM\, \frac{d\braket{\rho_\text{BH}(t_0)}}{dM},
\end{gather}
which is then compared with the constraints on a bin-by-bin basis. We find that for sufficient choices of the parameters $N_f$, $t_f$, $r_f$, and $r$, our model can produce black holes over practically the entire parameter space allowed by the constraints (see Fig. \ref{fig:CombinedConstraintPlots}).
Notably, this Figure illustrates two interesting points: 1) that this mechanism is capable of generating black holes which can account for both 100\% of the dark matter in the region $M \sim 10^{20}\text{ g}$ and production or r-process elements \cite{Fuller:2017uyd}, and 2) it is also capable of generating black holes with sufficient mass to explain the recent LIGO observation GW150914 \cite{Abbott16}. Some studies have even argued that PBH can account for 100\% of the DM in this range by contesting the CMB constraints \cite{Bird:2016dcv}. The three contours in Figure \ref{fig:CombinedConstraintPlots} are, however, simply chosen by hand for illustrative purposes, and are therefore not representative of the entire parameter space allowed to this mechanism, which is much wider than suggested by the given parameters.

\section{Parameter space}\label{sec:ParameterSpace}
\begin{figure}
\centering
\includegraphics[width=0.9\linewidth]{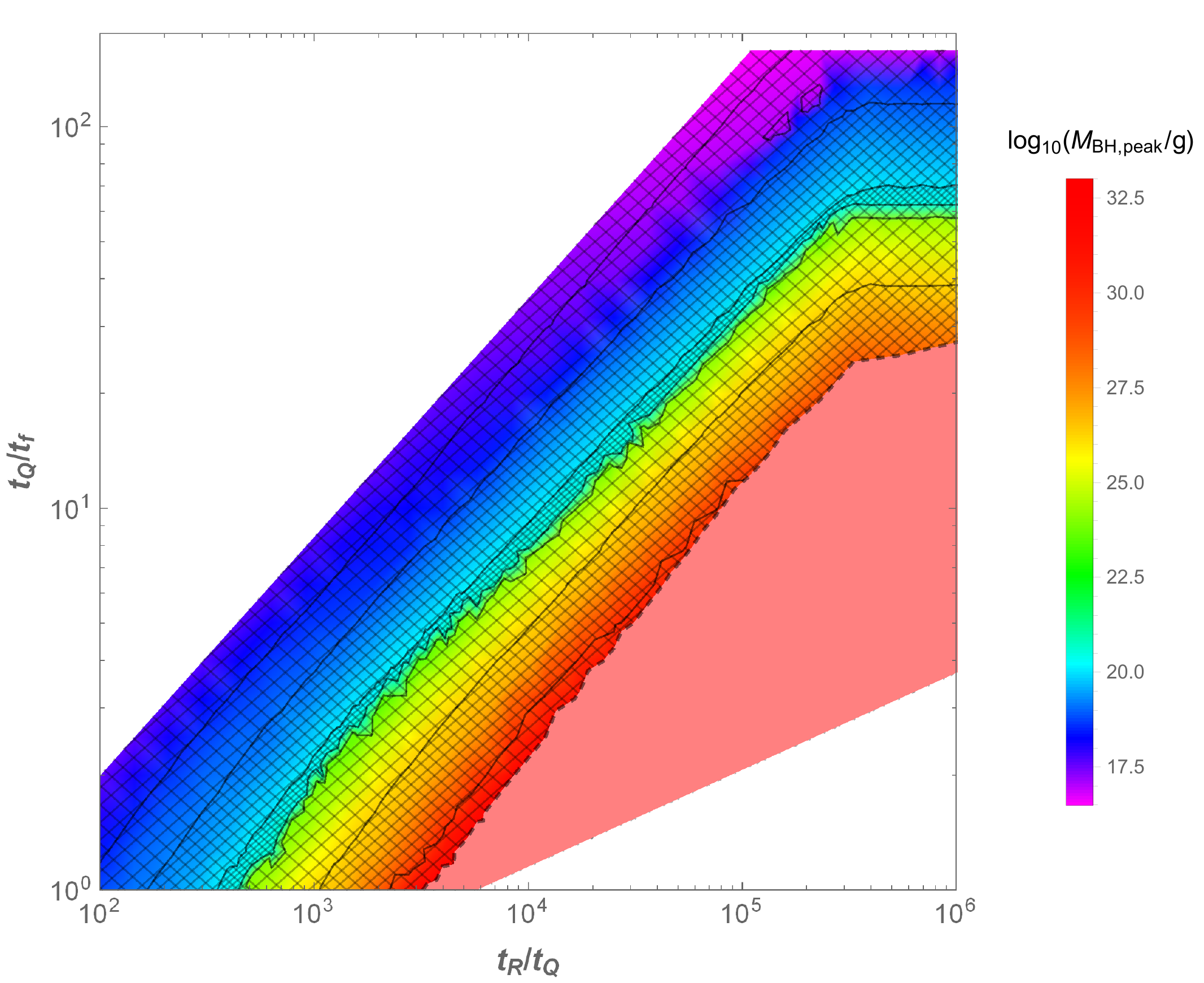}
\caption{Parameter space available to the model with $N_f = 10^6$. Color gradient denotes the peak black hole mass $M_\text{BH,peak}$ (warmer colors denote higher masses), and the black contours are surfaces of constant $f=\Omega_\text{PBH}/\Omega_\text{DM}$. Different levels of cross-hatching between $f$ contours indicate the range of maximum $f$ values, with the densest corresponding to $f \ge 1$ (highly correlated with $M_\text{BH,peak} = 10^{20}\text{ g}$), the next dense $10^{-1} < f < 1$, and so on, down to $f < 10^{-3}$. Pink shaded region to the bottom right is ruled out due to $T_R \ge 1\text{ MeV}$ for that set of parameters. Jagged edges of the contours are likely due to shortcomings of the interpolation method.}
\label{fig:ParameterSpace}
\end{figure}

We now explore the parameter space available to this mechanism in which it is possible to account for a considerable fraction of the dark matter while avoiding observational constraints. To do so, we develop an algorithm to accomplish this task in the following manner: since the function $((a_f/a_R)^3/\braket{\rho_Q(t_f)})\, (d\braket{\rho_\text{BH}(t_R)}/d\eta)$ is determined solely by the parameters $r_f$ and $r$, we generate a list of such functions by sampling the $r - r_f$ plane at various points. Then, for each $(r,r_f)$ pair, we vary $t_f$ using a weighed bisection method until $\max(|1-f_i/f_\text{con}(M_i)|) < \epsilon = 10^{-1}$, with $f_i$ given by Equation \ref{eq:DMfractionBin}. This determines the value of $t_f$ which gives the maximum dark matter fraction allowed by the constraints for the given values of $r$ and $r_f$. Once that has been determined, we can calculate all other relevant quantities of interest, such as $M_\text{BH,peak}$, $T_R$, and the dark matter fraction $f=\Omega_\text{PBH}/\Omega_\text{DM}$. The results for $N_f = 10^6$ are shown in Figure \ref{fig:ParameterSpace}.

We can see from this Figure that the contours of constant $f$ are highly correlated with the values of $M_\text{BH,peak}$. This is due to the fact that the observational constraints are solely a function of $M$, and the calculated spectrum $f(M)$ is quite sharply peaked at $M_\text{BH,peak}$. The region of $f \ge 1$ roughly follows the contour of $M_\text{BH,peak}=10^{20}\text{ g}$, where the constraints are weakest, due to this fact. The general tendency appears to be that for increasing $r$, the spectrum favors heavier black holes, while for increasing $r_f$, it favors lighter black holes. This is because the longer the period of structure formation, the more the large mass perturbations (which initially have small $\delta$) grow, increasing the probability that they collapse to black holes, whereas the longer the period between fragmentation and Q-ball domination, the more the energy density of Q-balls is diluted, so that the value of $t_f$ has to be lower in order to achieve the same value of $f$ that one would with smaller $r_f$. The smaller $t_f$, the smaller the horizon mass, and the smaller the PBH masses. This can be quantified, as the contours of constant $f$ very roughly follow $r^{0.63}/r_f \sim \text{const}$.

However, at around $r \gtrsim 3.2\times 10^{5}$, the character of the plot changes so that the spectrum appears to no longer depend on $r$, only on $r_f$. This is likely due to the definition of $\beta$ discussed in Section \ref{ssec:QballDensityPerturbations}, in which density perturbations are not allowed to grow past $\delta_c$, which makes it so that further increases in $r$ have no effect on the spectrum. Further increases in $r_f$, however, serve to further dilute the Q-ball density before structure formation can occur, lowering the necessary $t_f$ as mentioned above.

The pink region where $T_R > 1\text MeV$ is ruled out since in this region Q-ball decays begin to interfere with nucleosynthesis. We can see that it is correlated with high mass, as the later the fragmentation (meaning larger horizon mass), the less time there is for the Q-balls to dominate before the universe cools sufficiently enough that nucleosynthesis begins. One might notice that the mass of the black holes seen at LIGO in the GW150914 event lies beyond the range indicated in this Figure, but in Figure \ref{fig:CombinedConstraintPlots} we have plotted a contour that has $M_\text{BH,peak} \sim 30 \text{ M}_\odot$. This is because the contour in Fig. \ref{fig:CombinedConstraintPlots} is for $N_f = 10^5$, whereas Fig. \ref{fig:ParameterSpace} has $N_f$ fixed at $10^6$. We can see in Fig. \ref{fig:MassSpectra} that the density spectrum has a peak at higher masses for lower values of $N_f$, so that the equivalent plot to Fig. \ref{fig:ParameterSpace} for $N_f < 10^6$ would see the $T_R < 1 \text{ MeV}$ constraint pushed to higher values of $M_\text{BH,peak}$; enough so that they could explain the $30 \text{ M}_\odot$ black holes while avoiding the nucleosynthesis constraint.

\section{Topological defects}\label{sec:TopologicalDefects}
Topological defect formation can also lead to the production of PBHs if the topological defects come to dominate the energy density. The analysis is sufficiently different from that of Q-balls, primarily because typically only one defect per horizon is produced at the time of formation due to the Kibble mechanism~\cite{Kibble76}. However, the general mechanism remains the same: small number densities of defects lead to large fluctuations relative to the background density, these fluctuations become gravitationally bound and collapse to form black holes once the relic density has come to dominate, and the relics decay due to some instability (such as gravitational waves or decay to Nambu-Goldstone bosons in the case of cosmic strings). In order to accurately model production of PBHs from these defects, one should calculate the expected density perturbations on initially superhorizon scales, which only begin to grow once these scales pass back within the horizon and the defects come to dominate the universe's energy density.

Cosmic strings are probably the most likely candidate for primordial relics due to the fact that they are typically cosmologically safe, as the energy density in string loops is diluted during expansion at the same rate as radiation, $a^{-4}$ \cite{Vachaspati84,Turok1983}.  In contrast, the string ``network" (i.e. infinitely long strings) energy density redshifts as $a^{-2}$ so that they quickly come to dominate the universe's energy density. However, once long strings start intercommuting to produce loops, and these loops subsequently self-intersect to fragment into smaller loops, the string network approaches a scaling solution which leads to the $a^{-4}$ dilution of string loops~\cite{Bennett86}.

However, this scaling solution critically relies on the probability of string intercommutation being very close to unity so that the long strings can efficiently break into small loops. If this probability was sufficiently low, then the string density could redshift as $a^{-2}$ or $a^{-3}$, survive until a matter/string dominated era, initiate structure formation, collapse to form PBHs, and then subsequently decay. As long as these conditions are satisfied, cosmic strings could act as a source of PBH. In addition, there exists a large class of solutions to the string equations of motion which never self-intersect \cite{Kibble82}, making this scenario plausible.

\section{Discussion}\label{sec:Discussion}
The mechanism we have discussed has a number of advantages over some other models. It is extremely effective in creating primordial black holes across a broad range of masses, and it does not require the tuning the inflaton potential~\cite{Inomata:2017okj,GarciaBellido:1996qt,Inomata:2016rbd,Kawasaki:2016pql}. We did not have to make any ad-hoc assumptions regarding density fluctuations; the fluctuations are calculable from first principles.

The mechanism is also generally applicable to practically any complex scalar field with a conserved global charge and flat potential, so that the formation of PBH is now a general prediction of any theory containing such charged scalars. In particular, supersymmetric extensions to the Standard Model typically have such fields, making the production of primordial black holes a general prediction of such theories. For the case of supersymmetric Q-balls with the SUSY-breaking scale $\Lambda_\text{SUSY} > 10\text{ TeV}$, the fragmentation time cannot be much longer than the Hubble time $H^{-1} \sim M_p/g_*^{1/2} \Lambda_\text{SUSY} \lesssim 8 \times 10^{-15}\text{ s}$, which corresponds to peak PBH masses of about $10^{23} \text{g}$ (assuming $N_f \sim 10^6$). The solid black curve illustrated in Figure \ref{fig:CombinedConstraintPlots} satisfies this bound, thus primordial black holes from supersymmetric Q-balls can account for 100\% of the dark matter.

Supersymmetric Q-balls themselves have been suggested as the source of dark matter in models where they are entirely stable~\cite{Kusenko:2005du,Cotner:2016dhw,Kusenko:1997zq,Enqvist:1998xd,Kusenko:1997si}. However, the stability is model-dependent, and it only applies to Q-balls carrying baryonic charge (so-called B-balls), since those carrying leptonic charge (L-balls) would quickly evaporate to neutrinos \cite{Cohen:1986ct}. In our scenario, a short evaporation timescale is precisely what is needed to end the early era of Q-ball domination before nucleosynthesis begins. The L-balls would then be composed of slepton fields which subsequently decay to neutrinos at an early time. Since neutrinos do not decouple from the plasma until just before nucleosynthesis ($T \gtrsim \text{few MeV}$), they thermalize quickly. If they decay early enough, it may even be possible to generate the baryon asymmetry through conversion of lepton number to baryon number via sphaleron processes during the electroweak phase transition \cite{PhysRevD.64.123526}.

There are some remaining open questions, such as how well the assumption that all Q-balls within the volume $V$ are the same charge models this scenario. Simulations show that for scalar condensates with a high ratio of charge density to energy density, this is a good prediction, as all the Q-balls formed from this initial condition typically have similar sizes. This is also theoretically understood for a scalar condensate with a sufficiently large charge density~\cite{Kusenko:1997si,Enqvist:1997si,Dine:2003ax}. However, condensates with a large energy density and small charge density generally produce broad charge distributions, nearly symmetric about $Q=0$, since the excess energy cannot be contained in Q-balls with the same sign of $Q$ while also conserving charge. We suspect that in this scenario the production of PBH will be reduced, since the charge conservation does not play as significant a role. Loss of energy due to scalar radiation in the fragmentation process may still be able to produce energy density inhomogeneities, but this will require further study. In the same manner, the production of oscillons from the fragmentation of a real scalar field may be able to produce significant numbers of PBH as well.

One may also wonder what sort of mechanism is needed in order to ensure the Q-balls decay at the correct time. As an example, following the work of \cite{Cotner:2016dhw}, the lifetime of a Q-ball with initial charge $Q_0$ decaying to pseudo-Goldstone bosons through the effects of a charge-violating operator of the form $V_\slashed{Q}(\phi) = g \phi^n (\phi^*)^m/\Lambda_*^{n+m-4} + \text{c.c.}$ is given by
\begin{gather}
\tau \approx \frac{1-Q_0^{1-a}}{(a-1)\Gamma_0},
\end{gather}
where
\begin{gather}
a = \frac{1}{4}(7+2(n+m-2)),\\
\Gamma_0 = 112.7 |g|^2 e^{-0.236(n+m)} (n-m)^2 J_{nm} \Lambda(\Lambda/\Lambda_*)^{2(n+m)-8},
\end{gather}
and $J_{nm} \sim O(10^{-7} - 10^{-6})$. For $g \sim 0.1$, $\Lambda_* \sim 10^{16} \text{ GeV}$, and $\Lambda \sim 10^9 \text{ GeV}$, the lifetime of a Q-ball decaying through an operator with $(n,m) = (2,3)$ is about $\tau \sim 10^{-13} \text{ s}$, which is sufficient to explain the curve of Figure \ref{fig:CosmologicalTimeline} (and satisfies the SUSY bound). Decay through these higher-dimension operators isn't the only way to induce the decay of Q-balls though; many other scenarios have been explored in the literature \cite{Kusenko:2005du,Kawasaki:2005xc,Kasuya:2014ofa}.

This work also begs the question of what possible observables exist that could show the Q-ball clusters collapse to black holes. We assume that the collapse will produce a stochastic gravitational wave background \cite{PhysRevD.95.043511}, which could be detected by future observatories (or put constraints on the model). Further evolution of the PBH population could see successive mergers, which in addition to creating another stochastic GW background \cite{Garcia-Bellido:2017fdg}, could also alter the distribution of black hole masses (in addition to evaporation/accretion effects \cite{Bettwieser81,Custodio:1998pv}). We propose to calculate the gravitational wave spectrum in a future publication.

\section{Conclusion}\label{sec:Conclusion}
In summary, we have shown that the number density fluctuations of a Q-ball population in the early universe can lead to production of primordial black holes with sufficient abundance to explain the dark matter. Scalar fields and Q-ball formation are general features of supersymmetric extensions to the Standard Model, which provides a good motivation for this mechanism. A similar mechanism using solitons, topological defects, or other compact objects associated with scalar fields in the early universe can also lead to a copious production of primordial black holes.

\section{Acknowledgements}\label{sec:Acknowledgements}
This work was supported by the U.S. Department of Energy Grant No. DE-SC0009937.  A.K. was also supported by the World Premier International Research Center Initiative (WPI), MEXT, Japan.

\appendix
\section{Evolution of energy densities}\label{sec:EvolutionOfEnergyDensities}
Here we tabulate the functional form of the energy density for each species (radiation, Q-ball, black hole) up until the present day. The values for $t_f$, $t_Q$, $t_R$, and $\tau_Q$ are taken as input parameters (subject to some self-consistency conditions), while the values of $t_\text{eq}$ and $t_0$ are calculated from the procedure described in Section \ref{ssec:StandardCosmology}.

\subsection{Radiation}
The radiation density begins after reheating and is given by Equation \ref{eq:RadDensityRadDominated}. From this point we evolve it through time to the present day, taking into account the contribution due to Q-ball decays during the period $t_Q < t < t_R$.
\begin{widetext}
\begin{gather}
\rho_R(t) = \begin{cases}
\frac{\pi^2 M_p^2}{327 t^2} & t_\text{RH} < t < t_Q \\
\frac{\pi^2 M_p^2}{327 t_Q^2} \left[ 1 + \left(\frac{\tau_Q}{t}\right)^{2/3} \Gamma\left(\frac{5}{3}, \frac{t_Q}{\tau_Q}, \frac{t}{\tau_Q}\right) \right] \left(\frac{t_Q}{t}\right)^{8/3} & t_Q < t < t_R \\
\frac{\pi^2 M_p^2}{327 t_Q^2} \left[ 1 + \left(\frac{\tau_Q}{t_R}\right)^{2/3} \Gamma\left(\frac{5}{3}, \frac{t_Q}{\tau_Q}, \frac{t_R}{\tau_Q}\right) \right] \left(\frac{t_Q}{t_R}\right)^{8/3} \left(\frac{t_R}{t}\right)^2 & t_R < t < t_\text{eq} \\
\frac{\pi^2 M_p^2}{327 t_Q^2} \left[ 1 + \left(\frac{\tau_Q}{t_R}\right)^{2/3} \Gamma\left(\frac{5}{3}, \frac{t_Q}{\tau_Q}, \frac{t_R}{\tau_Q}\right) \right] \left(\frac{t_Q}{t_R}\right)^{8/3} \left(\frac{t_R}{t_\text{eq}}\right)^2 \left(\frac{t_\text{eq}}{t}\right)^{8/3} & t_\text{eq} < t < t_0
\end{cases}
\end{gather}

\subsection{Q-balls}
The Q-balls are created at the time of fragmentation $t_f$, and evolve as decaying nonrelativistic matter. The magnitude of the energy density becomes insignificant shortly after $t_R$. $M_f = \Lambda |Q_f|^\alpha N_f^{1-\alpha}$ can be determined from specifying $t_f$, $r_f$, $r$, and $N_f$, as given in Section \ref{sec:CosmologicalHistory}.
\begin{gather}
\braket{\rho_Q(t)} = \begin{cases}
\frac{3 M_f}{4\pi t_f^3} \left(\frac{t_f}{t}\right)^{3/2} e^{-(t-t_f)/\tau_Q} & t_f < t < t_Q \\
\frac{3 M_f}{4\pi t_f^3} \left(\frac{t_f}{t_Q}\right)^{3/2} \left(\frac{t_Q}{t}\right)^{2} e^{-(t-t_f)/\tau_Q} & t_Q < t < t_R \\
\frac{3 M_f}{4\pi t_f^3} \left(\frac{t_f}{t_Q}\right)^{3/2} \left(\frac{t_Q}{t_R}\right)^{2} \left(\frac{t_R}{t}\right)^{3/2} e^{-(t-t_f)/\tau_Q} & t_R < t < t_\text{eq} \\
\frac{3 M_f}{4\pi t_f^3} \left(\frac{t_f}{t_Q}\right)^{3/2} \left(\frac{t_Q}{t_R}\right)^{2} \left(\frac{t_R}{t_\text{eq}}\right)^{3/2} \left(\frac{t_\text{eq}}{t}\right)^{2} e^{-(t-t_f)/\tau_Q} & t_\text{eq} < t < t_0
\end{cases}
\end{gather}

\subsection{Black holes}
The black holes are created towards the end of the initial Q-ball dominated era, and their density at $t_R$ is given by Equation \ref{eq:BHdensity}:
\begin{align}
\braket{\rho_\text{BH}(t_R)} &= \left(\frac{a_f}{a_R}\right)^3 \sum_{N=1}^{\infty} \int_{V_\text{min}}^{V_R} \frac{dV}{V} \int_{0}^{\infty} dM\, \left(\beta \frac{M}{V}\right) F_Q \\
&= \left(\frac{t_f^{3/2} t_Q^{1/2}}{t_R^2}\right) \frac{M_f }{V_f} \sum_{N=1}^{\infty} \int_{x_\text{min}}^{x_R} dx\, \beta(x,N) x^{N+\alpha-2} e^{-x}
\end{align}
where
\begin{gather}
\beta(x,N) = \begin{cases}
\left(\left(\frac{N}{x}\right)^{1-\alpha} - 1\right) \left(\frac{t_R}{t_*}\right)^{2/3} \ge \delta_c : \begin{cases}
\gamma \delta_c^{13/2} \left(\frac{N}{N_f} \frac{x^\alpha}{N^\alpha}\right)^{13/3} r_f^{-13/2} & \beta \le 1 \\
1 & \beta > 1
\end{cases} \\
\left(\left(\frac{N}{x}\right)^{1-\alpha} - 1\right) \left(\frac{t_R}{t_*}\right)^{2/3} < \delta_c : \begin{cases}
\gamma \left(\left(\frac{N}{x}\right)^{1-\alpha} - 1\right)^{13/2} \left(\frac{t_R}{t_*} \frac{N}{N_f} \frac{x^\alpha}{N^\alpha}\right)^{13/3} r_f^{-13/2} & \beta \le 1 \\
1 & \beta > 1
\end{cases}
\end{cases}
\end{gather}
and
\begin{gather}
t_* = \begin{cases}
t_Q & x_\text{min} < x < x_Q \\
\frac{3 V_f x}{4\pi N_f t_f^{3/2} t_Q^{1/2}} & x_Q < x < x_R
\end{cases}
\end{gather}
where $x_\text{min} = N_f V_\text{min}/V_f = N_\text{min}$, $x_Q = N_f V_Q/V_f$, and $x_R = N_f V_R/V_f$. After this has been evaluated, the evolution of the black hole density is fairly straightforward:
\begin{gather}
\braket{\rho_\text{BH}(t)} = \begin{cases}
\braket{\rho_\text{BH}(t_R)} \left(\frac{t_R}{t}\right)^{3/2} & t_R < t < t_\text{eq} \\
\braket{\rho_\text{BH}(t_R)} \left(\frac{t_R}{t_\text{eq}}\right)^{3/2} \left(\frac{t_\text{eq}}{t}\right)^{2} & t_\text{eq} < t < t_0
\end{cases}
\end{gather}
\end{widetext}

\bibliography{pbh}
\end{document}